\documentclass[sigconf,balance=true,hidelinks,table]{acmart}

\usepackage{hyperref}
\usepackage{amsmath,amsfonts}
\usepackage{algorithmic}
\usepackage[ruled,linesnumbered,noend]{algorithm2e}
\usepackage{xcolor}
\usepackage{svg}
\usepackage{graphicx}
\usepackage{textcomp}
\usepackage{xcolor}
\usepackage{nccmath}
\usepackage{environ}
\usepackage{tcolorbox}
\usepackage{soul}
\usepackage{booktabs}
\usepackage{multirow}
\usepackage{array}
\usepackage{float}
\usepackage{makecell}
\usepackage{flushend}
\usepackage{enumitem}
\usepackage{hhline}
\usepackage{comment}
\usepackage{epstopdf} 
\usepackage{caption}
\usepackage{subcaption}

\usepackage{xcolor}
\usepackage{listings}

\usepackage{scalerel}

\NewEnviron{myequation}{%
\begin{equation}
\scalebox{1}{$\BODY$}
\end{equation}
}

\newcommand{\fig}{Fig.}
\newcommand{\tab}{Table}

\newcolumntype{P}[1]{>{\centering\arraybackslash}p{#1}}
\newcolumntype{R}[1]{>{\raggedleft\arraybackslash}p{#1}}

\newcommand{\code}[1]{\textnormal{\texttt{#1}}}

\usepackage{tikz}
\usetikzlibrary{shapes.arrows}
\newcommand*\circled[1]{\tikz[baseline=(char.base)]{
    \node[shape=circle,fill,inner sep=0.5pt] (char) {\textcolor{white}{#1}};}}

\definecolor{brightgreen}{rgb}{0.4, 1.0, 0.0}
\definecolor{ferrarired}{rgb}{1.0, 0.11, 0.0}

\newcommand{\UpArrow}{\begin{tikzpicture}[baseline=-0.3em]
\node[single arrow,draw,rotate=45,single arrow head extend=0.2em,inner
ysep=0.2em,transform shape,line width=0.0em,top color=brightgreen,bottom color=brightgreen] (X){};
\end{tikzpicture}}

\newcommand{\DownArrow}{\begin{tikzpicture}[baseline=-0.5em]
\node[single arrow,draw,rotate=-45,single arrow head extend=0.2em,inner
ysep=0.2em,transform shape,line width=0.0em,top color=ferrarired,bottom color=ferrarired] (X){};
\end{tikzpicture}}

\acmConference[ESEM 2024]{The 18th ACM/IEEE International Symposium on Empirical Software Engineering and Measurement}{20–25 October, 2024}{Barcelona, Spain}

\setcopyright{none}

\begin{document}

\title{Automatic Data Labeling for Software Vulnerability\\ Prediction Models: How Far Are We?}

\author{Triet Huynh Minh Le}
\affiliation{\institution{CREST - The Centre for Research on Engineering Software Technologies, The University of Adelaide}
\city{Adelaide}
\country{Australia}}
\affiliation{\institution{Cyber Security Cooperative Research Centre, Australia}
\city{}
\country{}}
\email{triet.h.le@adelaide.edu.au}

\author{M. Ali Babar}
\affiliation{\institution{CREST - The Centre for Research on Engineering Software Technologies, The University of Adelaide}
\city{Adelaide}
\country{Australia}}
\affiliation{\institution{Cyber Security Cooperative Research Centre, Australia}
\city{}
\country{}}
\email{ali.babar@adelaide.edu.au}

\begin{abstract}
\textbf{Background}: Software Vulnerability (SV) prediction needs large-sized and high-quality data to perform well.
Current SV datasets mostly require expensive labeling efforts by experts (human-labeled) and thus are limited in size.
Meanwhile, there are growing efforts in automatic SV labeling at scale. However, the fitness of auto-labeled data for SV prediction is still largely unknown.
\textbf{Aims}: We quantitatively and qualitatively study the quality and use of the state-of-the-art auto-labeled SV data, D2A, for SV prediction.
\textbf{Method}: Using multiple sources and manual validation, we curate clean SV data from human-labeled SV-fixing commits in two well-known projects for investigating the auto-labeled counterparts.
\textbf{Results}: We discover that 50+\% of the auto-labeled SVs are noisy (incorrectly labeled), and they hardly overlap with the publicly reported ones.
Yet, SV prediction models utilizing the noisy auto-labeled SVs can perform up to 22\% and 90\% better in Matthews Correlation Coefficient and Recall, respectively, than the original models.
We also reveal the promises and difficulties of applying noise-reduction methods for automatically addressing the noise in auto-labeled SV data to maximize the data utilization for SV prediction.
\textbf{Conclusions}: Our study informs the benefits and challenges of using auto-labeled SVs, paving the way for large-scale SV prediction.
\end{abstract}


\maketitle

\section{Introduction}

Software Vulnerabilities (SVs) like Log4Shell~\cite{log4j_vuln} have recently shaken software professionals.
This type of SV has the potential of resulting in colossal data breaches in millions of systems.
Ideally, these critical SVs should receive special attention and be detected and patched as soon as practical before attackers can exploit them.
In reality, manual SV detection needs substantial effort and expertise~\cite{le2021large,braz2022less}. The increasing complexity and size of software systems make it even more challenging for software and security practitioners to timely detect ever-growing SVs. These challenges have resulted in SVs staying hidden in codebases for years~\cite{meneely2013patch,le2021deepcva}, leaving affected software systems susceptible to dangerous exploits during the whole period. Thus, automated support for early SV detection is apparently needed to minimize the impacts of SVs and ease the burden on developers and security experts.

In the last decade, data-driven approaches like Machine Learning (ML) models have become popular for automated SV prediction/detection~\cite{ghaffarian2017software}. These models can benefit from steadily growing data of SVs in software/security repositories, i.e., surpassing 22k new ones in 2022~\cite{nvd_statistics}.
Existing SV prediction efforts (e.g.,~\cite{neuhaus2007predicting,scandariato2014predicting,ghaffarian2017software,jimenez2019importance,croft2021empirical,theisen2020better}) have mainly focused on developing various ML models to distinguish vulnerable code artifacts (e.g., files) from non-vulnerable ones.
Accordingly, these models demand suitable datasets containing both vulnerable and non-vulnerable code.

Currently, the development of SV prediction models has mostly relied on human-labeled SV data~\cite{hanif2021rise,croft2022data}.
The most common type of human-labeled data is in the form of manually reported Vulnerability-Fixing Commits (VFCs), i.e., code changes to fix SVs (e.g.,~\cite{fan2020ac,bhandari2021cvefixes}).
Affected/vulnerable code (e.g., files) can then be extracted from reported VFCs together with non-vulnerable code in non-VFCs to provide data for SV prediction.
Human-labeled SV data has been highly regarded and widely adopted as this type of data is explicitly acknowledged/vetted by software/security practitioners with real-world expertise and experience~\cite{croft2022data}.
However, manual reporting/labeling of SVs is resource-intensive and incomplete~\cite{sawadogo2022sspcatcher,zheng2021d2a}.
A significant number (60\%+) of VFCs have been reported missing in practice, e.g., due to silent fixes~\cite{nguyen2022hermes}, which implies that the actual number of VFCs is much more than what has been reported.
Importantly, these missing VFCs limit the size of most of the current human-labeled SV datasets and probably restrict the ability of ML models to effectively learn patterns from such data~\cite{croft2022data,lin2022vulnerability}.
These limitations, coupled with the usually higher criticality of SVs than other bugs, have motivated researchers to explore solutions to automatically label SVs/VFCs in the wild.

Automatic VFC/SV labeling\footnote{Our study mainly refers to human/automatic labeling of VFCs as human/automatic SV labeling unless specified otherwise.} aims to provide additional data in conjunction with human-labeled data for enhancing SV prediction, in the hope that ``\textit{more data beats a cleverer algorithm}''~\cite{domingos2012few}.
One such automatically labeled (auto-labeled) dataset, D2A, is increasingly trusted by researchers and practitioners to benchmark their SV prediction models (e.g.,~\cite{d2a_data,hanif2022vulberta,cheng2022bug,cheng2022path,steenhoek2023empirical}), sharing the vision of the renowned ImageNet dataset~\cite{russakovsky2015imagenet} in the Computer Vision domain.
Within the D2A framework, a commit is auto-labeled as a VFC if it satisfies two key criteria: (\textit{i}) it is fix/security-related and (\textit{ii}) it fixes at least one SV detected by a static analysis tool. The latter condition means that the detected SV only exists in the prior-fix version but not in the after-fix version of the current commit.
The D2A authors envisioned that this auto-labeled dataset has the potential to transform the field of data-driven SV prediction by alleviating the problem of limited data size.
We also found that the number of D2A-labeled SVs can be up to \sethlcolor{white}\hl{10} times larger than that of human-labeled SVs, as shown in Section~\ref{subsec:data_collection}.

There is, however, not yet a perfect SV labeling technique, and D2A is no exception.
Specifically, static analysis that forms the basis of D2A has been long known for producing many false positives/alarms~\cite{johnson2013don}, in turn adding serious noise to the auto-labeled data.
Thus, despite the benefits, the noisy nature of D2A raises concerns about the suitability of this data for SV prediction.
For example, it is unreliable to test SV prediction models on data with incorrectly labeled samples.
To the best of our knowledge, there is still a lack of systematic understanding about the quality of auto-labeled SV data like D2A as well as the impact of using such data for developing SV prediction models.

To bridge these gaps, our study investigates the utilization of auto-labeled data for SV prediction. Our \textbf{contributions} are three-fold.
\circled{1} We reveal quantitative and qualitative insights into the quality of \sethlcolor{white}\hl{3,391} auto-labeled SVs in the state-of-the-art D2A dataset~\cite{zheng2021d2a}, with respect to \sethlcolor{white}\hl{1,582} human-labeled SVs in the OpenSSL and FFmpeg projects.
\circled{2} We quantify the impact of using the \textit{large-sized yet potentially noisy} auto-labeled SVs on the performance of a wide range of SV prediction models.
The models target the file level, which balances between the quality and practicality of SV predictions~\cite{jimenez2019importance,croft2022noisy}.\footnote{The choice of file-level SV prediction is elaborated in Section~\ref{subsec:svp}.}
\circled{3} We explore noise-aware models based on noise-reduction techniques to tackle noisy auto-labeled data.
Overall, we provide researchers and practitioners with evidence-based knowledge about \textit{how much}, \textit{when}, and \textit{why} auto-labeled SVs can be effectively used for SV prediction. We also highlight areas for improving automatic data labeling for SV-related tasks.
We share the code and models at~\cite{reproduction_package_esem2024} to facilitate future research.

\noindent \textbf{Paper structure}. Section~\mbox{\ref{sec:background}} provides background on SV prediction and the required data. Section~\mbox{\ref{sec:rqs}} describes the three research questions. Section~\mbox{\ref{sec:method}} presents the methods used for answering the questions. Section~\mbox{\ref{sec:results}} analyzes our empirical results. Section~\mbox{\ref{subsec:threats_to_validity}} discusses the threats to validity. Section~\mbox{\ref{sec:conclusions}} concludes the study.

\section{Background and Motivation}\label{sec:background}

\subsection{Data-driven SV Prediction}
\label{subsec:svp}

Data-driven approaches have become a promising alternative to conventional static and dynamic analyses for code-based SV detection~\cite{ghaffarian2017software}.
Essentially, data-driven models leverage historical source code from project repositories to automatically learn patterns of SV and non-SV artifacts to distinguish them.
These models help minimize human effort and adapt better to new SVs as they do not rely on pre-defined detection rules or test oracles as in static and dynamic analyses, respectively~\cite{croft2021empirical}.

Data-driven SV approaches have been performed on various levels of granularity that serve different purposes. The granularity levels range from a whole project/package and software modules/files to functions and even individual code statements. The more fine-grained function and statement levels can reduce inspection effort for developers.
However, these levels may contain insufficient information for SV fixing (e.g., why the current function/statement is vulnerable)~\cite{croft2022data}.
Therefore, this study adopts the \textit{file} level as it has been widely used in the literature (e.g.,~\cite{croft2022noisy,neuhaus2007predicting,scandariato2014predicting,theisen2020better}) and in practice~\cite{morrison2015challenges,jimenez2019importance}.
The file level also generally induces fewer incorrectly labeled SVs than finer granularities~\cite{croft2022noisy}, which better aligns with our focus on SV data quality.

\begin{figure}[t]
    \centering
    \includegraphics[trim={16cm 0.8cm 13.5cm 0.2cm},clip,width=0.9\columnwidth,keepaspectratio]{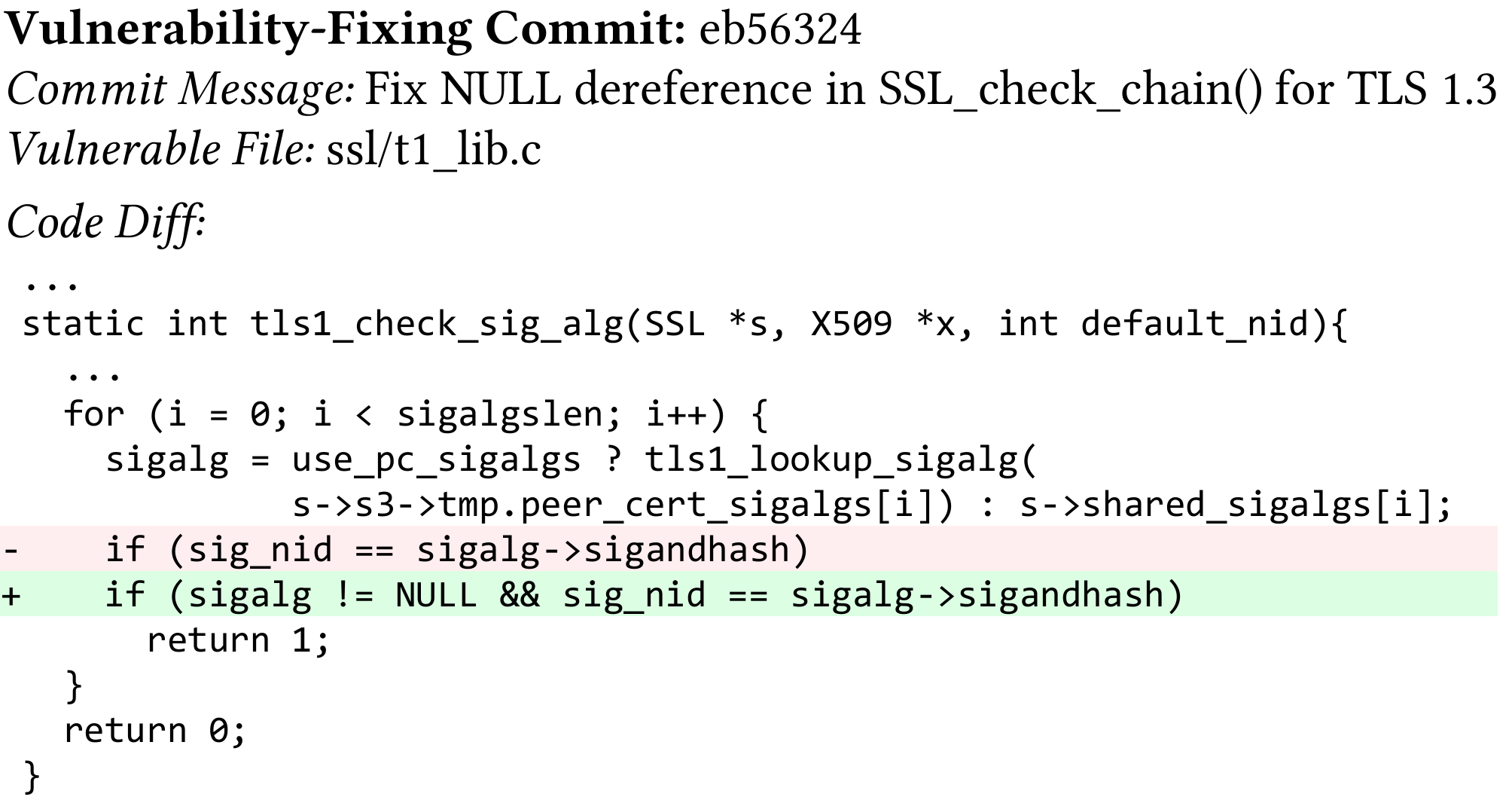}
    \caption{Exemplary vulnerable file corresponding to \textit{CVE-2020-1967} extracted from the respective vulnerability-fixing commit in the OpenSSL project.}
    \label{fig:file_ex}
\end{figure}

A data-driven model for file-level SV prediction requires an appropriate dataset of vulnerable and non-vulnerable files.
A file is usually considered vulnerable if it \textit{defines}, \textit{contains}, and/or \textit{uses} vulnerable code.
In \fig~\ref{fig:file_ex}, the exemplary vulnerable file \code{ssl/t1\_lib.c} contained an SV in the form of a null pointer dereference of the variable \code{sigalg}, leading to a crash if exploited.
On the other hand, a file is deemed non-vulnerable if it does not belong to any of the aforementioned categories.

\subsection{Data Labeling and Data Quality for SV Prediction Models}
\label{subsec:data_svp}

Currently, SV data are commonly collected through Vulnerability-Fixing Commits (VFCs) associated with SV reports.
We note that the non-SV data in a project are often the remaining artifacts that do not overlap with the collected SV data. Thus, it is important to reliably collect SV data to ensure the quality of the entire (SVs and non-SVs) dataset.
There are two main approaches to collecting VFCs in the wild: human labeling and automatic labeling. In this study, we work with real-world SVs, not synthetic ones like the Juliet Test Suite~\cite{juliet}. We also only consider labeling \textit{existing} SVs, not generating new/artificial SVs/defects~\cite{nong2022generating,kim2011dealing}.

\noindent \textbf{Human-labeled VFC/SV data}. Currently, human-labeled (validated by developers/experts) SV artifacts have been widely leveraged for SV prediction~\cite{hanif2021rise,croft2022data,le2024latent}.
After SVs are detected and reported, they are checked for validity/relevance to decide which SVs are worth fixing~\cite{panichella2021won}. The valid/relevant ones are fixed by project developers/maintainers.
Then, VFCs, i.e., commits that contain developers' changes (code deletions and additions) to fix the SVs, are checked and added to the SV reports.
Based on publicly reported VFCs in issue-tracking systems and/or SV databases like National Vulnerability Database (NVD)~\cite{nvd_website}, vulnerable artifacts (e.g., files) can be extracted.
Some existing SV datasets utilizing human-labeled VFCs are Big-Vul~\cite{fan2020ac}, CVEfixes~\cite{bhandari2021cvefixes}, and CrossVul~\cite{nikitopoulos2021crossvul}.
We follow the existing practice to extract prior-fix code from VFCs as vulnerable, e.g., the file \code{ssl/t1\_lib.c} with the vulnerable line \code{if (sig\_nid == sigalg->sigandhash)} in the VFC \textit{eb56324} in \fig~\ref{fig:file_ex}.

\noindent \textbf{Auto-labeled VFC/SV data}.
Recently, automatic labeling of VFCs is becoming an emerging research direction because a non-negligible number (60+\%) of existing VFCs have not been explicitly labeled as such in practice~\cite{nguyen2022hermes}.
Such missing VFCs are mainly due to \textit{silent fixes}; i.e., developers commit changes to fix SVs but do not label/report the commits as VFCs~\cite{sabetta2018practical}.

Earlier efforts (e.g.,~\cite{sabetta2018practical,zhou2017automated,chen2020machine}) have relied on pre-defined keywords in commit messages to retrieve missing VFCs. However, this keyword-based approach tends to generate many false positives because a security word can have a non-security meaning. For instance, the ``\textit{hash}'' keyword is a technique in cryptography but also a data structure in programming. To reduce such false positives when predicting VFCs, commit code changes can be used in combination with commit messages.

D2A~\cite{zheng2021d2a} is the state-of-the-art auto-labeled SV dataset generated by a VFC labeling technique that utilizes both commit messages and code changes~\cite{lin2022vulnerability}.
D2A is increasingly used because it provides not only VFCs but also useful details such as types and locations of detected SVs for fixing. The D2A technique has three main steps: (\textit{i}) selecting commits whose messages contain fix/security-relevant keywords, (\textit{ii}) performing a differential SV analysis on the prior-fix and after-fix versions of the selected commits using a static analysis tool, and (\textit{iii}) refining obtained VFCs (e.g., removing duplicate/irrelevant ones).
If a commit only passes the first step (e.g., no SVs detected or SVs persisting after the fix), then it would be considered a non-VFC by D2A.
Among these steps, the key novelty/contribution of D2A is the use of the Infer~\cite{infer_tool} static analyzer to identify SVs that disappear after a fix is applied, indicating that the current commit is likely to be a VFC. The SV differential analysis is automatically applied to the whole history of a project, so D2A has the ability to uncover missing VFCs and enlarge existing SV datasets.

The use of static analysis in D2A, however, leads to possible noise in the auto-labeled SV data and in turn raises concerns about the use of the data for SV prediction. Static analysis is commonly known for generating a large number of false positives, i.e., non-VFCs labeled as VFCs~\cite{croft2021empirical,johnson2013don}.\footnote{Section~\ref{subsec:rq1_results} demonstrates that many false positives indeed exist in the D2A dataset.}
Despite the noise caused by such false positives, most studies (e.g.,~\cite{song2022hgvul,hanif2022vulberta,cheng2022bug,cheng2022path,steenhoek2023empirical}) have used D2A ``\textit{as-is},'' even for evaluating SV prediction models.
To the best of our knowledge, there has been no systematic quality validation of (D2A) auto-labeled SVs used for SV prediction.
There is also little known about the extent to which the auto-labeled SV data overlap and/or complement conventional human-labeled SV data, though both data types have been used to develop SV prediction models.

Data quality for SV prediction has gained traction in the field~\cite{croft2023data}. Latent SVs in human-labeled data is the key issue that many recent studies (e.g.,~\cite{jimenez2019importance,croft2022noisy,garg2022learning,le2024latent}) have pointed out.
The extraction of (non-)vulnerable code from (non-)VFCs can also be inaccurate, e.g., due to tangled changes~\cite{herzig2013impact}.
We have leveraged existing recommendations and (manual) validation to build the ``\textit{golden}'' ground-truth data (see Section~\ref{subsec:data_collection}), but our focus is fundamentally different from theirs.
To the best of our knowledge, we are the first to investigate the quality of auto-labeled VFC/SV data as well as the impact and the ways of using such data for (file-level) SV prediction models, with respect to the conventional human-labeled counterparts.

\begin{figure*}[t]
    \centering
    \includegraphics[width=0.95\textwidth,keepaspectratio]{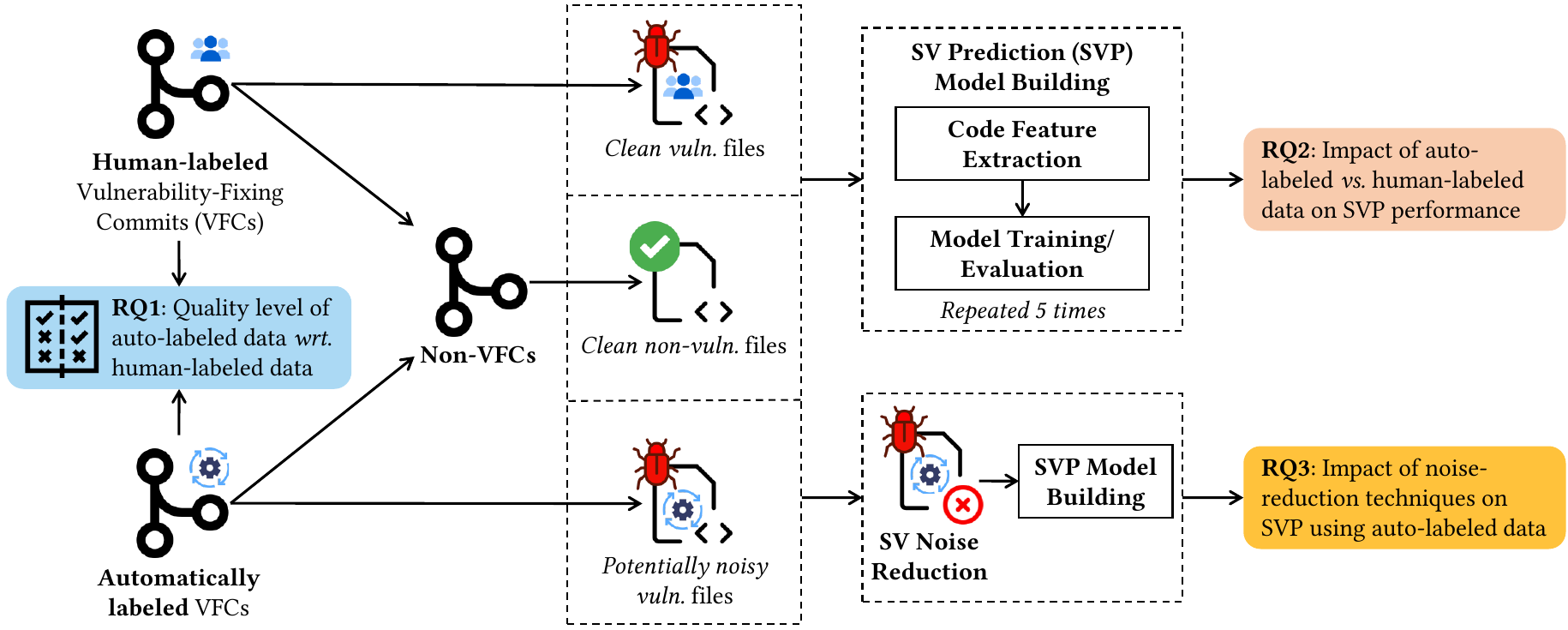}
    \caption{Research methods for answering the three research questions. \textbf{Note}: Non-VFCs are the commits not fixing SVs.}
    \label{fig:methodology}
\end{figure*}

\section{Research Questions}
\label{sec:rqs}

We answer three Research Questions (RQs) to demystify the quality of auto-labeled VFCs/SVs and their utilization for SV prediction.

\textbf{RQ1: What is the quality level of auto-labeled SVs with respect to human-labeled SVs?}
Given that human-labeled and auto-labeled VFC/SV data can both be used for SV prediction, RQ1 compares them to explore how much and why they (dis-)agree. Such comparison can distill systematic knowledge about the quality (noise level) of auto-labeled SV data. Human-labeled SV data are used as the baseline for the comparison as they are commonly expert-vetted and most widely used for SV prediction, as demonstrated in Section~\ref{subsec:data_svp}.
The findings of RQ1 are expected to help raise researchers' and practitioners' awareness of the potential benefits and threats when using auto-labeled SV data for SV prediction.

\textbf{RQ2: To what extent do auto-labeled SVs contribute to SV predictive performance?}
There have been separate uses of human-labeled and auto-labeled SV data for SV prediction, yet there has been no direct comparison between them.
RQ2 compares the performance of SV prediction models using auto-labeled SVs (as is), human-labeled SVs, and both together.
RQ2 findings would inform how much auto-labeled SVs can complement conventional human-labeled SVs to improve SV predictive performance.

\textbf{RQ3: Do noise-reduction techniques improve the performance of SV prediction models using auto-labeled SVs?}
Auto-labeled SV data are inevitably noisy (see Section~\ref{subsec:data_svp}), but manual removal of such noise is nearly impossible due to huge manual effort. RQ3 explores different types of noise-reduction techniques aiming at automatically tackling such noise and improving SV predictive performance in RQ2.
In RQ3, we do not propose novel noise-reduction techniques. Instead, we highlight how effective the noise-reduction techniques currently used in Software Engineering and related domains can handle the noise in auto-labeled SV data.

\section{Case Study Design}
\label{sec:method}

\noindent \fig~\ref{fig:methodology} illustrates the research methods we used to answer the three RQs targeting file-level SV prediction, as explained in Section~\ref{subsec:svp}.

\noindent \textbf{Overview}.
The case study design had three key components: (\textit{i}) collection of vulnerable files from human-labeled/auto-labeled VFCs and non-vulnerable files from non-VFCs (section~\ref{subsec:data_collection}), (\textit{ii}) development of SV prediction models (sections~\ref{subsec:code_features},~\ref{subsec:svp_algorithms}, and~\ref{subsec:model_evaluation}), and (\textit{iii}) use of noise-reduction techniques for SV prediction (section~\ref{subsec:noise_reduction}).
RQ-wise methods based on the components are described hereafter.

\noindent \textbf{RQ1}. We analyzed overlapping and divergent cases between all the curated human-labeled and auto-labeled VFCs/SVs.
The divergent cases enabled us to unveil why human-labeled SVs were missed by auto-labeling and vice versa.
Such analysis would reveal the (noisy) nature of auto-labeled SVs as the human-labeled data were validated.
We manually validated the labels of a random set of significant size with 90\% confidence and 10\% error~\cite{cochran2007sampling} of \textit{auto-labeled but not human-labeled} SVs.
For labeling, we first read the commit message, code changes, and linked bug report (if any) to comprehend each commit.
To reduce subjectivity, we only labeled a commit as VFC if an SV fix was evident to us in either the commit message, code changes, or a linked public SV report.
For example, in the OpenSSL project used in our study (see Section~\ref{subsec:data_collection}), the commit \textit{59a56c4} contained ``Add NULL check'' in the message, and the check was added in line 73; thus, we labeled it as VFC.
We documented the reasoning behind our labels and then performed thematic analysis~\cite{braun2006using} of the labeling reasoning to identify the patterns of the false positives of auto-labeling.
The analysis was done by the first author with 3+ years of experience in the SV and Software Engineering areas. The results were checked by the second author, an expert in Software Engineering and Software Security with 20+ years of experience. Any disagreements were resolved through discussions.
Details of our manual analysis can be found at~\cite{reproduction_package_esem2024}.

\noindent \textbf{RQ2}. We used the obtained vulnerable and non-vulnerable files to develop three types of SV prediction models. The types were using vulnerable files that are (\textit{i}) human-labeled only, (\textit{ii}) auto-labeled only, and (\textit{iii}) both human-labeled and auto-labeled.
To provide compatible input for the models, different methods were utilized to extract code features from the vulnerable and non-vulnerable files. Such features were then fed into various Machine Learning (ML) algorithms to distinguish the vulnerable files from the non-vulnerable ones. The process of feature extraction and model training/evaluation was repeated five times to improve the stability of results.

\noindent \textbf{RQ3}. We investigated various noise-reduction techniques to mitigate the noise (false positives) in the auto-labeled vulnerable files found in RQ1, aiming to improve the performance of the models using such data in RQ2.
These techniques removed auto-labeled SVs that were deemed greatly different from human-labeled ones or likely to result in a model predicting non-SVs.
The resultant noise-aware models utilized the same feature extractors and the training/evaluation procedures in RQ2. To decipher the inner workings of the noise-aware models, like RQ1, we also analyzed the auto-labeled samples removed and retained by such models.

\subsection{Data Collection}
\label{subsec:data_collection}

This section presents the curation of human-labeled and auto-labeled vulnerable files as well as non-vulnerable files. These files are needed to assess the quality of auto-labeled SVs and their impact on SV prediction models in comparison with human-labeled SVs.
We targeted C/C++ for data collection as this language has been widely studied~\cite{croft2022data} and adopted in practice~\cite{popular_languages}.
Specifically, we selected the \textit{OpenSSL}\footnote{\url{https://github.com/openssl/openssl}} and \textit{FFmpeg}\footnote{\url{https://github.com/FFmpeg/FFmpeg}} projects for data collection because of the following three reasons:

\begin{itemize}
    \item OpenSSL and FFmpeg are long-term open-source projects with thousands of active contributors and mature SV reporting processes. The frequent contributions enhance the quality of input code for developing SV prediction models. The high-standard SV reporting processes help maximize the number of human-labeled SVs. More human-labeled SVs increase the reliability of using such data as the baseline comparison with auto-labeled SVs, which is the focus of this study. Publicly available repositories and data of these projects also support the reproducibility of the research.
    \item The two projects have been used ubiquitously. OpenSSL provides the important SSL and TLS protocols for 60+\% of the websites~\cite{jimenez2019importance}. FFmpeg is one of the most popular multimedia (audio/video) processing libraries nowadays and has been integrated into Google products like YouTube. Given their popularity, SVs in these projects like Heartbleed~\cite{heartbleed} would probably result in catastrophic impacts worldwide, making OpenSSL and FFmpeg highly relevant to our study.
    \item There is a large number of auto-labeled (D2A) SVs in these two projects, which is essential for the focused investigations in this study.
    We did not use the other projects having D2A-labeled data as they had too limited ($<$ 20) human-labeled VFCs to compare with the auto-labeled ones.
\end{itemize}

We collected vulnerable files from VFCs and then non-vulnerable files from Non-VFCs.
The VFCs were labeled manually by practitioners/experts or automatically by the D2A technique~\cite{zheng2021d2a}.
The vulnerable files from human-labeled VFCs were called human-labeled as they were validated (see Section~\ref{subsubsec:human_labeled_svs}).
The non-VFCs fixed bugs that we considered to be non-SV, but were not always human-labeled (in the commit message) as \textit{not fixing an SV}.
Thus, we did not call the non-vulnerable files human-labeled to avoid confusion.
The numbers of the collected files are given in \tab~\ref{tab:data_statistics}.

\begin{table}[t]
\fontsize{8}{9}\selectfont

  \centering
  \caption{The number of human-labeled \& auto-labeled vulnerable files as well as non-vulnerable files from the OpenSSL \& FFmpeg projects used in this study.}
 \begin{tabular}{llll}
    \hline
    \textbf{Data type} & \textbf{OpenSSL} & \textbf{FFmpeg} & \textbf{Total} \\
    \hline
    Human-labeled vulnerable files & 141 & 1,441 & 1,582 \\
    Auto-labeled (D2A) vulnerable files & 1,733 & 1,658 & 3,391 \\
    Non-vulnerable files & 18,787 & 54,255 & 73,042 \\
    \hline
    \end{tabular}
  \label{tab:data_statistics}
\end{table}

\subsubsection{\textbf{Extraction of human-labeled vulnerable files}}\label{subsubsec:human_labeled_svs}
We collected human-labeled VFCs from OpenSSL and FFmpeg for extracting human-labeled vulnerable files.
While many existing studies (e.g.,~\cite{jimenez2019importance,fan2020ac,bhandari2021cvefixes,croft2021empirical}) have mainly used NVD~\cite{nvd_website} for collecting VFCs, our study augmented NVD with the \textit{official} security advisories that are frequently updated by the maintainers of OpenSSL\footnote{\url{www.openssl.org/news/vulnerabilities.html}} and FFmpeg.\footnote{\url{www.ffmpeg.org/security.html}}
To the best of our knowledge, the use of these advisories is the first time in the literature.
We only considered human-labeled VFCs publicly reported (on NVD and the advisories) as it is impractical to manually search for all VFCs.
We obtained \sethlcolor{white}\hl{121} and \sethlcolor{white}\hl{1,385} \textit{unique} VFCs corresponding to \sethlcolor{white}\hl{85} and \sethlcolor{white}\hl{343} SVs in OpenSSL and FFmpeg, respectively. 
These numbers imply that an SV could be fixed in more than one VFC. We still treated each VFC independently as it is unrealistic to know in advance whether a VFC is a partial fix for the SV in real-world scenarios.
Notably, \sethlcolor{white}\hl{46} (38\%) and \sethlcolor{white}\hl{1,260} (91\%) of these VFCs came from the security advisories of OpenSSL and FFmpeg, respectively. These significant increases in size demonstrate the value of the security advisories in obtaining VFCs.

From these VFCs, we extracted \sethlcolor{white}\hl{1,604} \textit{candidate} vulnerable files (159 in OpenSSL and 1,445 in FFmpeg), i.e., the prior-fix versions of the affected files in the commits.
These files were obtained after we removed test files to focus on production code and discarded files with only cosmetic (non-functional) changes, e.g., changing whitespaces/newlines/comments. These filtering steps are common practices in the literature (e.g.,~\cite{croft2021empirical,croft2022noisy,li2021vulnerability,wattanakriengkrai2020predicting}).
We did not trace/include latent vulnerable files as there is not yet an accurate way to automatically determine the origin (introduction time) of SVs~\cite{croft2022noisy}.

To ensure the quality of human-labeled vulnerable files, we validated and removed non-SV files from the above candidates. Specifically, \sethlcolor{white}\hl{1,409} files (63 in OpenSSL and 1,346 in FFmpeg) were the only file modified in their respective VFCs and thus had to be vulnerable. For the remaining \sethlcolor{white}\hl{195} vulnerable files belonging to VFCs with 2+ affected files, we manually checked them and discarded \sethlcolor{white}\hl{22} ones unlikely to be vulnerable (not defining/containing/using vulnerable code).
This was done by the first author and validated by the second author with conflicts resolved in discussions. The validation took an extensive effort of 90 man-hours.
We found two key reasons for the false positives. The first scenario was the files adding a whole new function to fix an SV, e.g., the file \code{libavcodec/pthread.c} in the VFC \textit{59a4b73} of FFmpeg. These added functions are necessary for SV fixing, but they can be defined in other files. The current file is a placeholder for the SV-fixing code and thus not vulnerable.
Another scenario was the files only containing new configurations/flags used elsewhere to fix SVs. For instance, the file \code{ssl/ssl\_err.c} in the VFC \textit{b77ab01} of OpenSSL defined a new error message for tackling CVE-2016-2181 in the file \code{ssl/d1\_pkt.c} of the same commit.
Notably, very few ($\approx$1\%) irrelevant cases also imply that tangled changes in human-labeled VFCs in the OpenSSL and FFmpeg projects are unlikely to impact SV predictive performance.
After validation, we got \textit{141} and \textit{1,441} vulnerable files from the human-labeled VFCs for OpenSSL and FFmpeg, respectively.

\subsubsection{\textbf{Extraction of auto-labeled vulnerable files}}\label{subsubsec:auto_labeled_svs}
We collected \sethlcolor{white}\hl{441} and \sethlcolor{white}\hl{1,144} \textit{unique} VFCs auto-labeled by D2A~\cite{zheng2021d2a} for OpenSSL and FFmpeg, respectively. The D2A authors mentioned that it might take 12+ hours to statically analyze a single code version, which made it impossible for us to rerun the differential analysis for all the commits in OpenSSL and FFmpeg within a reasonable amount of time. Thus, we used the published VFCs of D2A~\cite{d2a_data}.
We followed the same filtering practices described in Section~\ref{subsubsec:human_labeled_svs} to extract vulnerable files from the D2A-labeled VFCs.
We also refer to the D2A-labeled vulnerable files as auto-labeled data.
We retrieved \textit{1,733} and \textit{1,658} D2A vulnerable files for OpenSSL and FFmpeg, respectively.
Among them, \sethlcolor{white}\hl{789} in OpenSSL and \sethlcolor{white}\hl{1,260} in FFmpeg were included in the traces of SV reports of the static analyzer~\cite{infer_tool} used by D2A.
These trace-included files were more likely to be vulnerable than the others in the D2A-labeled VFCs. We evaluated both file variants for SV prediction.
We note that there was no guarantee of the label correctness of the D2A vulnerable files because the security relevance of the commits was not confirmed. Manual validation of all D2A-labeled SVs was also not affordable due to its sheer size, i.e., 17+ times larger than what we manually validated for the human-labeled ones. We still analyzed a subset of the D2A files in Section~\ref{subsec:rq1_results}.

\subsubsection{\textbf{Extraction of non-vulnerable files}}\label{subsubsec:non_svs}
We extracted non-vulnerable files from the remaining commits of OpenSSL and FFmpeg that were not human-labeled or D2A-labeled as VFCs.
Using the extensive keyword lists from~\cite{pletea2014security,le2020puminer}, we removed commits with messages containing any listed security words as these commits might be SV-related.
There is also an increasing concern about \textit{latent} SVs inducing false-negative data for prediction models~\cite{croft2022data,croft2022noisy}. Latent SVs are mainly due to either partially fixed SVs or SVs existing in prior versions/commits of a fixed vulnerable file in a VFC. To improve the quality of non-vulnerable files, we discarded all the non-vulnerable files having the same name or content as those in the collected VFCs. Although this is seemingly a conservative approach, it has been shown to be effective in removing latent SVs~\cite{croft2021empirical}.
We then examined a significant sample of 68 non-vulnerable files~\cite{cochran2007sampling} in each project but did not observe any false negatives.
Despite the validation results, we assert that non-vulnerable files cannot be perfectly clean in practice given the current lack of perfect security testing.
Finally, we obtained \textit{18,787} and \textit{54,255} non-vulnerable files for OpenSSL and FFmpeg, respectively.

\subsection{Code Feature Extractors}
\label{subsec:code_features}

Raw data of code files entered six popular feature extractors to produce compatible inputs for SV prediction models.
The feature types were: Bag-of-Tokens (similar to Bag-of-Words with code tokens as words), Bag-of-Subtokens (character sequences of code tokens), Word2vec~\mbox{\cite{mikolov2013distributed}}, fastText~\mbox{\cite{bojanowski2017enriching}}, Doc2vec~\cite{le2014distributed}, and CodeBERT~\mbox{\cite{feng2020codebert}}. These features have been widely used for SV prediction (e.g.,~\cite{jimenez2019importance,croft2022data,kalouptsoglou2021empirical,hin2022linevd,fu2022linevul,le2024software}).
We did not use software metrics as they have been shown to underperform the above text-mining ones~\cite{le2020deep,croft2021empirical,scandariato2014predicting}.
For all the features except CodeBERT, we used a code-aware tokenizer to capture code semantics/syntax. For example, \code{a-{}-} is split into \code{a} and \code{-{}-} to explicitly inform a model that the variable \code{a} is decreased by one. For Bag-of-Subtokens and fastText, subtokens had lengths from two to six. We note that one is too noisy and a length of more than six is likely to explode the vocabulary size and computational cost. In addition, Word2vec and fastText produce token-wise vectors; thus, the vector of a file was the average of the feature vectors of all the constituent tokens in the file.
For CodeBERT, we reused its pre-trained model because it had been custom-made for code-related tasks. We used the vector of the \textit{[CLS]} token to represent each file, as recommended in the original CodeBERT work. For any files with 512+ tokens (the limit of input length for CodeBERT), we split the file into multiple blocks with a maximum size of 512 tokens each and averaged the vectors of these blocks.

\subsection{SV Prediction Algorithms}
\label{subsec:svp_algorithms}

We applied six ML algorithms to leverage the extracted features to classify vulnerable files from non-vulnerable ones. The classifiers were: K-Nearest Neighbors (KNN), Support Vector Machine (SVM), Logistic Regression (LR), Random Forest (RF), Light Gradient Boosting Machine (LGBM)~\mbox{\cite{ke2017lightgbm}}, and XGBoost (XGB)~\mbox{\cite{chen2016xgboost}}.
To optimize the classifiers, we performed a grid search of the following hyperparameters:
KNN: \textit{no. of neighbors}: \{11, 31, 51\}, \textit{distance weight}: \{uniform, distance\}, and \textit{distance norm}: \{1, 2\};
SVM and LR: \textit{regularization coefficient}: \{0.01, 0.1, 1, 10, 100\} and \textit{regularization norm}: \{1, 2\};
RF, LGBM, and XGB: \textit{no. of estimators}: \{100, 300, 500\}, \textit{max. no. of leaf nodes}: \{100, 200, 300, unlimited\}, and \textit{max. depth}: \{3, 5, 7, 9, unlimited\}.
The selected classifiers and hyperparameters have been adopted for SV-related tasks (e.g.,~\mbox{\cite{le2020puminer,spanos2018multi,le2019automated,le2022survey,duan2021automated,le2022towards,le2022use,nguyen2024automated}}).
We focused on ML as it is currently commonly used for file-level SV prediction (e.g.,~\cite{ghaffarian2017software,jimenez2019importance,croft2022data,theisen2020better}).

\subsection{Model Evaluation}
\label{subsec:model_evaluation}

\subsubsection{\textbf{Evaluation technique}}\label{subsubsec:evaluation_technique}
\begin{figure}[t]
    \centering
    \includegraphics[width=0.97\columnwidth,keepaspectratio]{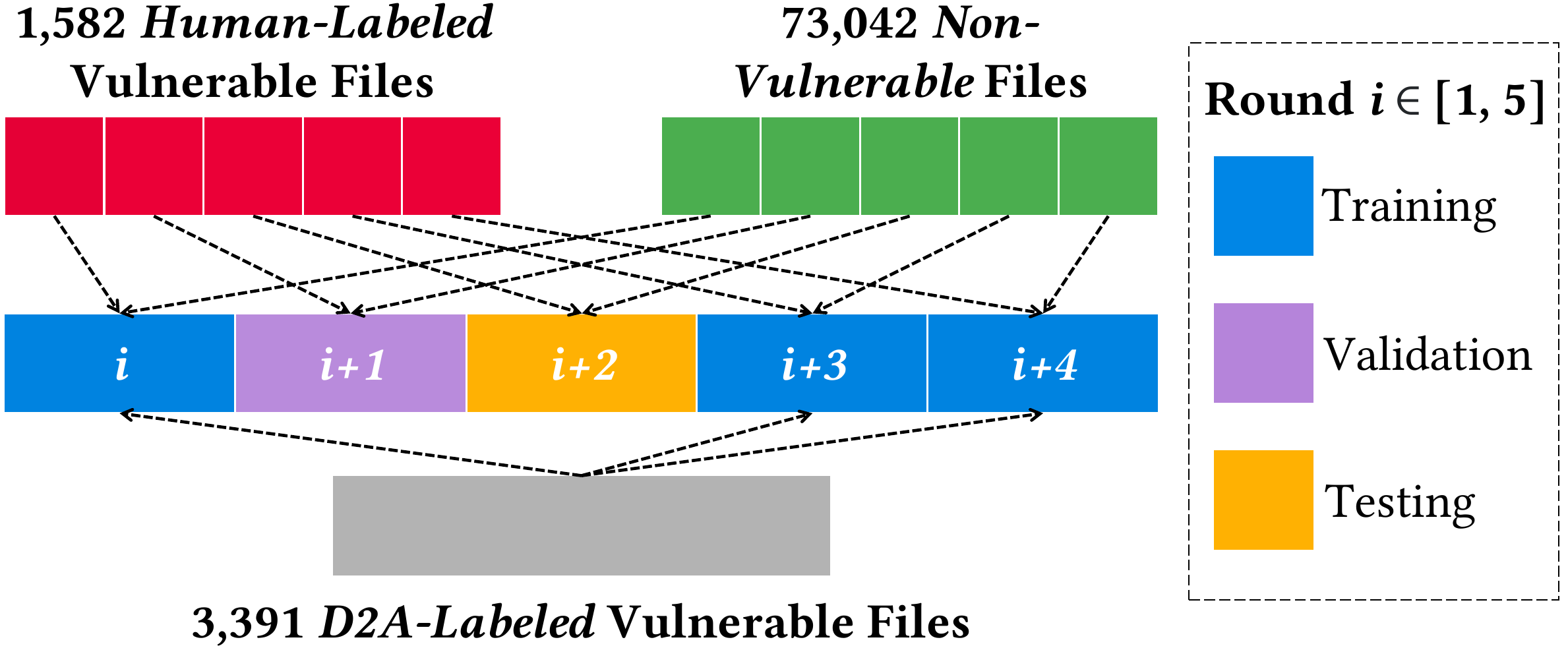}
    \caption{The 5-round training, validation, \& testing file-level SV prediction models. \textbf{Notes}: The splits are of equal size. Any index exceeding five would be wrapped around (e.g., 6\%5 = 1).}
    \label{fig:evaluation_splits}
\end{figure}

We used five rounds of training, validation (hyperparameter tuning), and testing to train and evaluate file-level SV prediction models, as illustrated in \fig~\ref{fig:evaluation_splits}. This evaluation technique struck a balance between the reliability of results and the requirement of computational resources.
Particularly, from the human-labeled vulnerable files and non-vulnerable files, we generated five randomly stratified data splits. These splits had an equal ratio of vulnerable to non-vulnerable files. Stratification has been shown to decrease biases during evaluation compared to the traditional k-fold cross-validation~\cite{wattanakriengkrai2020predicting}.
In each round, we used all D2A-labeled vulnerable files (excluding the duplicates with the human-labeled vulnerable files if any) for training but not for validation/testing the models as these cases were not validated (see Section~\ref{subsubsec:auto_labeled_svs}).
It is worth noting that release-based evaluation~\cite{jimenez2019importance} was not possible in this study because some of the releases containing the human-labeled vulnerable files did not have any D2A-labeled vulnerable files for evaluation.
We also did not apply any class rebalancing techniques like random over-sampling to avoid their additional effects when evaluating the performance of SV prediction with auto-labeled data compared to that of human-labeled data.
The use of auto-labeled and human-labeled data with class rebalancing techniques can be explored in the future.

\subsubsection{\textbf{Evaluation measures}}\label{subsubsec:evaluation_measures}
We used Precision, Recall, F1-Score, and Matthews Correlation Coefficient (MCC) to measure the performance of file-level SV prediction models.
The ranges for the first three measures and MCC are [0, 1] and [-1, 1], respectively, where 1 is the best value.
These measures have been previously used for SV prediction (e.g.,~\cite{jimenez2019importance,croft2021empirical,croft2022noisy,theisen2020better}).
We used MCC for selecting optimal models because MCC performs evaluation with both classes~\mbox{\cite{luque2019impact}}.
For RQ2 and RQ3, we recorded the \textit{testing} performance of the optimal models with the highest validation MCC.

\subsection{Noise-Reduction Techniques}
\label{subsec:noise_reduction}

With the aim of improving SV predictive performance, we investigated three techniques for automatically tackling noisy auto-labeled vulnerable files (i.e., files labeled as vulnerable but actually non-vulnerable).
The noise-reduction techniques were: (\textit{i}) Confident Learning~\cite{northcutt2021confident}, (\textit{ii}) Centroid-based Removal, and (\textit{iii}) Domain-specific Removal.
The noise-aware models with the cleaned data used the same feature extractors (section~\ref{subsec:code_features}) and ML-based classifiers (section~\ref{subsec:svp_algorithms}).
We did not apply noise-reduction methods to human-labeled data as it is beyond the scope of the study and we validated such data in sections~\ref{subsubsec:human_labeled_svs} and~\ref{subsubsec:non_svs}. Without the ground-truth labels of auto-labeled SVs (see Section~\ref{subsubsec:auto_labeled_svs}), we could only report the impact of noise-reduction on the downstream SV predictive performance, but not noise-reduction performance itself.

\noindent \textbf{Confident Learning (CL)}. CL~\cite{northcutt2021confident} is the state-of-the-art method for removing data noise in many domains such as audio/text processing, and recently Software Engineering~\cite{eken2021deployment,paleyes2022challenges}.
To ``clean'' noisy data samples, CL needs the labels/classes and the class-wise predictive probabilities generated by a trained model for each of the samples. The noise level is estimated as the number of samples labeled as vulnerable but with a predictive probability for the non-vulnerable class equal to or higher than the average non-vulnerable probability ($t_{non\_vuln.}$).
Note that $t_{non\_vuln.}$ is the average probability of predicting the files in the non-vulnerable class as non-vulnerable. Based on the noise level ($k$), $k$ noisy samples are selected from SV-labeled files with the highest predictive probabilities for the non-vulnerable class.
We did not directly use models trained on human-labeled data for reducing noise in auto-labeled data without CL (i.e., removing the D2A files not predicted as vulnerable) as CL was originally shown to outperform this strategy~\cite{northcutt2021confident}.
In our study, CL identified and removed noisy D2A-labeled files using the models learned from the human-labeled vulnerable files and non-vulnerable files in the training set of each evaluation round.

\noindent \textbf{Centroid-based Removal (CR)}. Unlike the model-dependent CL, CR is a model-agnostic technique for detecting noisy samples. CR is adapted from Positive Unlabeled learning commonly used in the Software Engineering domain (e.g.,~\cite{le2020puminer,tian2012identifying}). CR computes the cosine distances between a sample to the \textit{centroids} of the vulnerable and non-vulnerable classes. The centroid of a class is the average of the feature vectors of all the data samples in that class. CR removes the samples that are SV-labeled yet are closer (having a smaller distance) to the centroid of the non-vulnerable class.
In our study, CR computed the distances between each D2A-labeled vulnerable file to the centroids of human-labeled vulnerable files and non-vulnerable files in the training set of each evaluation round.

\noindent \textbf{Domain-specific Removal (DR)}. While CL and CR are the general noise-reduction methods for various tasks, DR is tailored to filter (potentially) noisy samples in the D2A dataset. DR removes the files in the D2A-labeled VFCs that were not included in the traces of SV reported generated by Infer~\cite{infer_tool}, the static analysis tool used by D2A. As mentioned in Section~\ref{subsubsec:auto_labeled_svs}, these files are less likely to be affected by the detected SVs and in turn probably be noisier.

\section{Experimental Results and Discussion}
\label{sec:results}

\subsection{\textbf{RQ1}: What Is the Quality Level of Auto-labeled SVs With Respect To Human-labeled SVs?}
\label{subsec:rq1_results}

\definecolor{LeftColor}{HTML}{ABD9F4}
\definecolor{MiddleColor}{HTML}{4BAE4F}
\definecolor{RightColor}{HTML}{FFC23B}

\begin{figure}[t]
    \centering
    \includegraphics[trim={15.5cm 0.3cm 15cm 0.3cm},clip,width=\columnwidth,keepaspectratio]{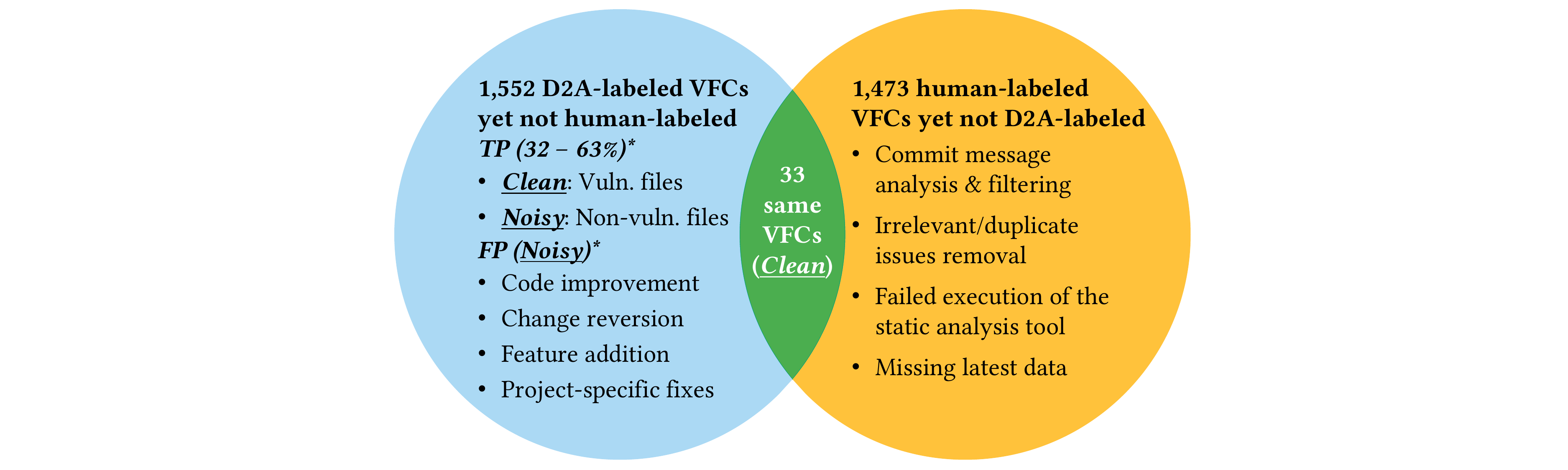}
    \caption{The relationship between the auto-labeled (D2A) SVs \& human-labeled SVs. \textbf{Notes}: (*) indicates that the results were obtained from a subset of 68 samples from each of the OpenSSL \& FFmpeg projects. The overlapping VFCs were only from the FFmpeg project.}
    \label{fig:ald_hld}
\end{figure}

\sethlcolor{MiddleColor}\hl{\mbox{\textcolor{white}{\textbf{Same VFCs}}}}. Despite sharing the goal of labeling VFCs, only $\approx$2\% (33/1,585) of all the D2A-labeled VFCs overlapped with the human-labeled ones (see \fig~\mbox{\ref{fig:ald_hld})}.
All of these overlaps were in FFmpeg.
The human-labeled and D2A types of SVs mostly agreed, except for the four cases related to \textit{null pointer dereference}. For example, the VFC \textit{b829da36} in FFmpeg was tagged with \textit{buffer overflow (CWE-119)} by human labeling and \textit{null pointer dereference (CWE-476)} by D2A, while the reversed case occurred for the VFC \textit{837cb43}. We also found that the files in 31/33 (94\%) of these VFCs were all included in the SV traces of D2A. These results show that D2A can identify SVs in these overlapping cases with high accuracy.

\noindent \sethlcolor{RightColor}\hl{\textbf{Human-labeled yet not D2A-labeled VFCs}}.
We identified the key causes of 1,473 (121 in OpenSSL and 1,352 in FFmpeg) such VFCs.
There were 148 cases (28 in OpenSSL and 120 in FFmpeg) published after the D2A commits had been collected.\footnote{Last commit dates: 24 Sep 2019 (OpenSSL) and 18 Apr 2020 (FFmpeg)} For the 1,325 human-labeled VFCs on the same date or before the D2A VFCs, only 47 of them (all in FFmpeg) were deemed non-vulnerable by D2A. This finding seemingly suggested that the commit message analyzer~\cite{d2a_cma} of D2A might have filtered out these missing VFCs (i.e., for not containing the pre-defined keywords) because any commits included in the D2A dataset needed to first pass the commit message filtering (see Section~\ref{subsec:data_svp}). Interestingly, when we reran the commit message analyzer using the default configurations, only 88 (34 in OpenSSL and 54 in FFmpeg) cases were not selected by the analyzer. We speculate that the remaining 1,190 human-labeled VFCs were not included in the D2A dataset due to other reasons, e.g., removal of duplicate/irrelevant issues or failed execution of the static analysis tool. The exact analysis of these cases was not possible because we could not rerun the differential static analysis due to the lack of computational resources, as mentioned in Section~\ref{subsubsec:auto_labeled_svs}.

\noindent \sethlcolor{LeftColor}\hl{\textbf{D2A-labeled yet not human-labeled VFCs}}.
We manually analyzed 68 cases labeled vulnerable by D2A but not reported on NVD/security advisories in each project.
We found 22/68 (32\%) in OpenSSL and 43/68 (63\%) in FFmpeg were vulnerable with SV fixes mentioned in the commit messages.
These numbers suggest that 50+\% of auto/D2A-labeled VFCs in the two projects are potentially noisy.
Our analysis also showed that files included in SV traces by the static analyzer used by D2A could still be non-vulnerable and vice versa.
Hereafter, we discuss the true and false positives of D2A VFCs to improve understanding of the potential benefits and challenges when using such auto-labeled data for SV prediction, which has never been studied in the literature.
We do not discuss true/false negatives of D2A as we focus on auto-labeled SVs.

\noindent \textbf{True-positive D2A VFCs}. The SV types of these cases were either same or a sub-class of those human-labeled.
For example, D2A assigned the SV fixed in the VFC \textit{7ab6312} in FFmpeg with CWE-195, a child of CWE-681 in the human-labeled VFCs.
In some of these cases, developers explicitly acknowledged the existence of an SV in the commit, yet mentioned why the SV was not publicly reported with a CVE. For example, in the commit \textit{a3e9d5a} in OpenSSL, the committer reported that a proof-of-concept side-channel attack had been demonstrated for the vulnerable code. However, such an attack only affected the localhost, so the SV would have minimal impact and was not given a CVE. Despite the limited impact, these SVs/VFCs are perhaps still of value as the ``\textit{Local}'' Attack Vector is recognized by the widely used Common Vulnerability Scoring System~\cite{cvss_spec}.
Furthermore, D2A had VFCs not yet reported on either NVD or the security advisories, but their respective fixed SVs were already reported on NVD.
An example of such silent fixes is the missing commit \textit{c046fff} in OpenSSL that fixed CVE-2002-0659.
These cases show that D2A contains relevant VFCs/SVs that can complement the human-labeled ones on NVD or security advisories.

\noindent \textbf{False-positive D2A VFCs}.
These False Positive (FP) commits were not acknowledged as VFCs by developers/experts.
They contained D2A keywords that have both general and SV meanings (e.g., \textit{issue} or \textit{error}); their code changes were also incorrectly flagged by the static analyzer used by D2A.
We used thematic analysis~\cite{braun2006using} to distill patterns of these FPs. The patterns are: (\textit{i}) Existing code improvement, (\textit{ii}) Existing change reversion/removal, (\textit{iii}) New feature addition, or (\textit{iv}) Project-specific fixes.

\textit{Existing code improvement}. This category mostly improves different aspects of an application without adding new functionality to the original code. The commits in this category enhance four key quality attributes: Maintainability, Usability, Performance, and Compatibility. The improvements are mainly made either by changing code structure/organization (e.g., code refactoring) or switching existing algorithms/code to a newer/improved version. For example, the commit \textit{9156a5a} in FFmpeg modified how the variables were defined to make them more intuitive (i.e., changing from video offsets to actual coordinates), which in turn helped improve the maintainability/understandability of the code.

\begin{figure*}[t]
    \begin{subfigure}{\textwidth}
         \centering
         \includegraphics[width=\textwidth,keepaspectratio]{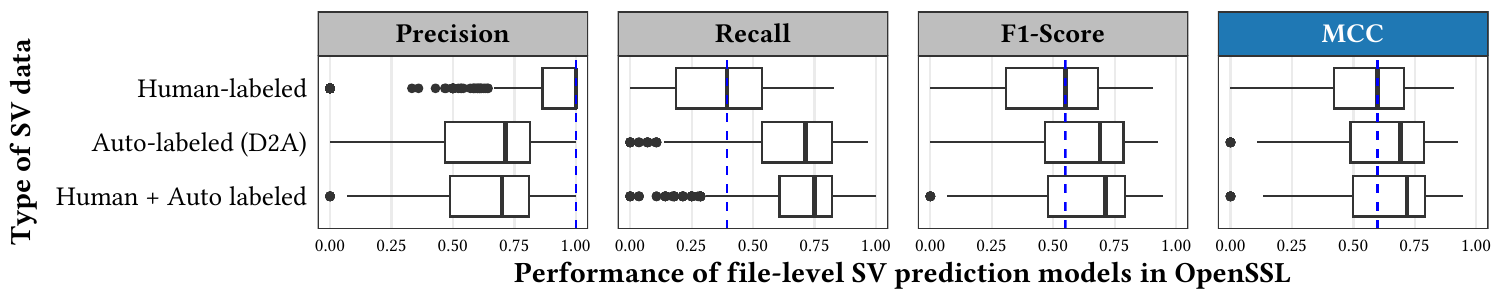}
         \label{fig:openssl_performance}
         \vspace{-20pt}
     \end{subfigure}
     \rule{\textwidth}{0.01pt}
     \begin{subfigure}{\textwidth}
         \centering
         \includegraphics[width=\textwidth,keepaspectratio]{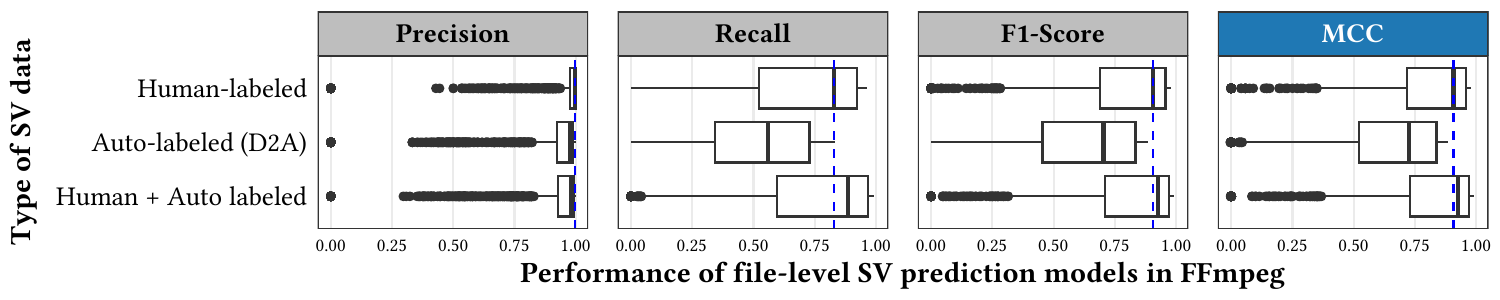}
         \label{fig:ffmpeg_performance}
     \end{subfigure}
     \caption{Testing performance of file-level SV prediction models using the three data types. \textbf{Notes}: The blue vertical dashed lines are the median performance values of the models using only human-labeled SVs. MCC is the main evaluation measure.}
    \label{fig:data_type_comparison}
\end{figure*}

\textit{Existing change reversion/removal}. This category mainly disables or removes a functionality/feature/support added prior to the current commit. Such a reversion/removal is commonly done to support compatibility with other modules/components or fix a related issue that was discovered after the original change had been made. For instance, the commit \textit{bd990e2} in OpenSSL removed SSL/TLS fragmented alerts. Such change increased the compatibility when communicating with other components, given that some of these components no longer support this type of alerts.

\textit{New feature addition}.
This category adds a new feature/functionality/support as part of the continuous integration process~\cite{arani2024systematic} to an application. Such an addition is performed to enhance the capability of the application and/or serve a new use case.
For instance, the commit \textit{2986ecd} in OpenSSL enabled the existing function \code{EVP\_PKEY\_copy\_parameters()} to work with non-provided (i.e., \code{NULL}) parameters. This added support would be particularly useful when reusing content from an existing entity.

\textit{Project-specific fixes}. This category fixes errors involving application logic. These fixes often require an understanding of the requirements/operations of the application. For instance, the commit \textit{a6191d0} in FFmpeg reversed the order of the data layout to prevent data copies during Huffman encoding, which was an error specific to the FFmpeg application.

Besides the false-positive VFCs, the noise in D2A also comes from non-vulnerable files in real VFCs, i.e., \textit{tangled changes}~\cite{herzig2013impact}. The average numbers of files per D2A-labeled commit were \sethlcolor{white}\hl{4.19 and 1.49} in OpenSSL and FFmpeg, respectively, higher than those (\sethlcolor{white}\hl{1.44 and 1.05}) of the human-labeled VFCs.
These statistics tend to imply the existence of tangled commits as intuitively, changes are made in many places/files to achieve multiple purposes.
To verify this hypothesis, we examined the large D2A commits affecting more than five files each. Indeed, we found true-positive VFCs that contained tangled changes.
For example, a non-SV change, i.e., removal of duplicate code, was made in the file \code{crypto/x509v3/v3\_ncons.c}, a part of the VFC \textit{776654a} in OpenSSL that mainly fixed an out-of-memory SV.
There were also D2A files that only added whole functions or configurations to fix SVs, which were SV-related but not vulnerable themselves. For instance, the file \code{libavformat/utils.c} in the VFC \textit{4641ae3} in FFmpeg added a function to perform a check to prevent an out-of-memory SV.
Overall, the noise patterns of auto-labeled SVs involve complex code semantics, which is non-trivial to remove using manually defined rules.
Thus, we explore automatic noise-reduction techniques for tackling such noise in Section~\ref{subsec:rq3_results}.

\begin{tcolorbox}
\textbf{RQ1 Summary}.
Auto-labeled SVs (by D2A) are large-sized, but they are mostly misaligned with human-labeled SVs.
Auto-labeled data contain relevant SVs in silently reported VFCs.
Yet, a significant portion (50+\%) of the D2A-labeled SVs are potentially noisy. The noise is mainly caused by missing security keywords in the commit message analyzer, the sub-optimal performance of the static analysis tool, and tangled commits.
The noise suggests that auto-labeled data are not reliable to be used for \textit{testing} SV prediction models.
\end{tcolorbox}

\begin{table}[t]
\fontsize{8}{9}\selectfont
  \centering
  \caption{Average \& best (in parentheses) testing performance of file-level SV prediction models using the three data types. \textbf{Notes}: Best/optimal models have the highest validation MCC. Bold values are the highest row-wise. Gray cells have the highest average \& best values in each row.}
    \begin{tabular}{lccc}
    \hline
    \multirowcell{3}[0ex][l]{\textbf{Evaluation}\\ \textbf{measure}} & \multicolumn{3}{c}{\textbf{Data type}}\\
    \cline{2-4}
    & \makecell{\textbf{Human-labeled} \\ \textbf{data (HLD)}} & \makecell{\textbf{Auto-labeled}\\ \textbf{data (D2A)}} & \makecell{\textbf{Combined data} \\ \textbf{(HLD + D2A)}} \\
    \hline
    \multicolumn{4}{c}{\cellcolor[HTML]{9AD2F2} \textbf{OpenSSL}}\\
    \hline
    Precision & \cellcolor[HTML]{C0C0C0}\textbf{0.833} (\textbf{0.905}) & 0.613 (0.858) & 0.618 (0.853) \\
    Recall & 0.374 (0.695) & 0.657 (0.845) & \cellcolor[HTML]{C0C0C0}\textbf{0.712} (\textbf{0.851})\\
    F1-Score & 0.531 (0.799) & 0.621 (0.851) & \cellcolor[HTML]{C0C0C0}\textbf{0.650} (\textbf{0.851})\\
    MCC & 0.530 (0.797) & 0.618 (0.850) & \cellcolor[HTML]{C0C0C0}\textbf{0.647} (\textbf{0.850})\\
    \hline
    \multicolumn{4}{c}{\cellcolor[HTML]{9AD2F2} \textbf{FFmpeg}}\\
    \hline
    Precision & \cellcolor[HTML]{C0C0C0}\textbf{0.914} (\textbf{0.996}) & 0.881 (0.979) & 0.888 (0.985) \\
    Recall & 0.699 (0.942) & 0.510 (0.774) & \cellcolor[HTML]{C0C0C0}\textbf{0.747} (\textbf{0.977})\\
    F1-Score & 0.787 (0.970) & 0.653 (0.870) & \cellcolor[HTML]{C0C0C0}\textbf{0.805} (\textbf{0.981})\\
    MCC & 0.784 (0.969) & 0.649 (0.868) & \cellcolor[HTML]{C0C0C0}\textbf{0.801} (\textbf{0.981})\\
    \hline
    \multicolumn{4}{c}{\cellcolor[HTML]{9AD2F2} \textbf{Average of the projects}}\\
    \hline
    Precision & \cellcolor[HTML]{C0C0C0}\textbf{0.874} (\textbf{0.951}) & 0.747 (0.918) & 0.753 (0.919)\\
    Recall & 0.532 (0.819) & 0.584 (0.810) & \cellcolor[HTML]{C0C0C0}\textbf{0.729} (\textbf{0.914}) \\
    F1-Score & 0.661 (0.885) & 0.637 (0.861) & \cellcolor[HTML]{C0C0C0}\textbf{0.727} (\textbf{0.916}) \\
    MCC & 0.659 (0.883) & 0.633 (0.859) & \cellcolor[HTML]{C0C0C0}\textbf{0.724} (\textbf{0.916}) \\
    \hline
    \end{tabular}%
  \label{tab:avg_best_data_type}%
\end{table}%

\subsection{\textbf{RQ2}: To What Extent Do Auto-labeled SVs Contribute to SV Predictive Performance?}
\label{subsec:rq2_results}

The models trained with \textit{only} SVs auto-labeled by D2A (D2A models) had reasonable performance.
\tab~\ref{tab:avg_best_data_type} and \fig~\ref{fig:data_type_comparison} shows that F1-Score and MCC of the D2A models were 0.6+ on average and 0.85+ at best, which can be useful for developers~\cite{neuhaus2007predicting,shin2013can}.
The results also confirmed that less noise in auto-labeled data enhanced model performance. RQ1 showed that D2A files in FFmpeg were less noisy than those in OpenSSL, so the D2A models in FFmpeg outperformed those in OpenSSL by 5.2\% F1-Score and 5\% MCC.
The promising performance shows that the true-positive vulnerable files in (D2A) auto-labeled data can benefit the \textit{training} of SV prediction models as a form of SV data augmentation~\cite{le2024latent,le2024mitigating}.
Yet, it is important to note that the auto-labeled data contain noisy/false-positive samples, and thus they are unreliable/untrustworthy for validating/testing the models.

The D2A models performed on par and even beat the models using the clean Human-Labeled Data (HLD models).
Compared to the HLD models, the D2A models were 16.9\% (F1-Score) and 16.6\% (MCC) better for OpenSSL, but 17\% (F1-Score) and 17.2\% (MCC) worse for FFmpeg.
These findings can be explained by the true positive rates of D2A-labeled SVs in each project. Extrapolating the rates (32\% in OpenSSL and 63\% in FFmpeg) in RQ1 would roughly lead to 0.32 $\times$ 1,733 = 555 true-positive SVs ($\approx$4 times more than HLD) in OpenSSL and 0.63 $\times$ 1,658 = 1,045 true-positive SVs (similar size to HLD) in FFmpeg. This implies that auto-labeled data are particularly useful when HLD are limited. For OpenSSL, the auto-labeled data started to beat the human-labeled data when the former was $\approx$4 times the size of the latter, but had 51\%$\downarrow$ in MCC with equal size.
This suggests that if only SVs/VFCs from NVD had been used as mostly done in the literature, then the gains of the D2A models over the HLD models would have been even more evident.
When the two data types are of similar size, HLD tend to produce better model performance because of the clean nature.
This ``cleanliness'' is reinforced by the higher Precision of the HLD models than that of the D2A models.
Contrarily, the D2A models tended to have competitive to strong Recall values because auto-labeled data can provide more diverse/unseen yet noisier SV patterns. For example, RQ1 showed that D2A provided patterns of SVs that have been long missing in the human-labeled dataset (e.g., CVE-2002-0659).

Combining HLD and auto-labeled (D2A) data had the best performance (see \tab~\ref{tab:avg_best_data_type} and \fig~\ref{fig:data_type_comparison}).
The combined models, on average, produced 9.9\% (up to 22.1\% for OpenSSL) and 14.4\% (up to 23.4\% for FFmpeg) higher MCC than the HLD and D2A models, respectively.
The two average improvements were confirmed statistically significant by the Wilcoxon signed-rank tests~\mbox{\cite{wilcoxon1992individual}} with $p$-values of 1.2e-4 and 2.4e-5 $<$ 0.01 and medium to large effect sizes of 0.429 and 0.611, respectively.\footnote{Effect size $(r) = Z / \sqrt{N}$, where $Z$ is the test statistic, $N$ is the total no. of samples~\cite{wattanakriengkrai2020predicting}; $r \leq 0.1$: negligible, $0.1 < r \leq 0.3$: small, $0.3 < r \leq 0.5$: medium, $r > 0.5$: large~\cite{tomczak2014need}.}
Similar improvements were observed for the other measures, except Precision. The outperformance was also observed for both projects, with up to 90\% increase in Recall recorded for OpenSSL.
Note that the results of each ML algorithm can be found at~\cite{reproduction_package_esem2024}; algorithm-wise results still follow the overall patterns.

\begin{tcolorbox}
\textbf{RQ2 Summary}. Auto-labeled SVs are noisy yet can improve SV predictive performance (up to 22+\%$\uparrow$ in MCC).
The improvements are substantial when auto-labeled SVs are of a much (4$\times$) larger size than human-labeled SVs.
We recommend combining auto-labeled with human-labeled SVs for training models to maximize performance gains.
\end{tcolorbox}

\begin{table}[t]
\fontsize{6.9}{7.9}\selectfont
  \centering
  \caption{Percentage (\%) differences in average testing performance between the models in RQ2 \& the noise-aware models using the three noise-reduction techniques in RQ3: Confident Learning (CL)~\cite{northcutt2021confident}, Centroid-based Removal (CR), \& Domain-specific Removal (DR). \textbf{Note}: HLD is Human-Labeled Data.}
    \begin{tabular}{lR{0.8cm}R{0.8cm}R{0.8cm}|R{0.8cm}R{0.8cm}R{0.8cm}}
    \hline
    \multirowcell{3}[0ex][l]{\textbf{Evaluation}\\ \textbf{measure}} & \multicolumn{6}{c}{\textbf{Noise-aware models}}\\
    \cline{2-7}
    & \multicolumn{3}{c|}{\textbf{Auto-labeled (D2A)}} & \multicolumn{3}{c}{\textbf{Combined (HLD + D2A)}}\\
    \cline{2-7}
    & \textbf{CL} & \textbf{CR} & \textbf{DR} & \textbf{CL} & \textbf{CR} & \textbf{DR}\\
    \hline
    \multicolumn{7}{c}{\cellcolor[HTML]{9AD2F2} \textbf{OpenSSL}}\\
    \hline
    Precision & 31.45 \UpArrow & 20.57 \UpArrow & 18.00 \UpArrow & 28.83 \UpArrow & 19.62 \UpArrow & 17.05 \UpArrow\\
    Recall & --17.5 \DownArrow & --16.9 \DownArrow & --17.0 \DownArrow & --14.6 \DownArrow & --9.06 \DownArrow & --10.1 \DownArrow\\
    F1-Score & 6.752 \UpArrow & 0.445 \UpArrow & --1.29 \DownArrow & 7.782 \UpArrow & 4.903 \UpArrow & 2.786 \UpArrow\\
    MCC & 6.971 \UpArrow & 0.569 \UpArrow & --1.17 \DownArrow & 7.980 \UpArrow & 5.033 \UpArrow & 2.924 \UpArrow\\
    \hline
    \multicolumn{7}{c}{\cellcolor[HTML]{9AD2F2} \textbf{FFmpeg}}\\
    \hline
    Precision & 4.779 \UpArrow & 1.546 \UpArrow & 0.180 \UpArrow & 4.973 \UpArrow & 0.946 \UpArrow & 0.709 \UpArrow\\
    Recall & --1.16 \DownArrow & --4.33 \DownArrow & --4.86 \DownArrow & --0.65 \DownArrow & --0.11 \DownArrow & --0.61 \DownArrow\\
    F1-Score & 2.349 \UpArrow & --1.01 \DownArrow & --2.51 \DownArrow & 2.770 \UpArrow & 0.526 \UpArrow & --0.04 \DownArrow\\
    MCC & 2.358 \UpArrow & --1.04 \DownArrow & --2.52 \DownArrow & 2.783 \UpArrow & 0.534 \UpArrow & --0.03 \DownArrow\\
    \hline
    \multicolumn{7}{c}{\cellcolor[HTML]{9AD2F2} \textbf{Average of the projects}}\\
    \hline
    Precision & 18.11 \UpArrow & 11.06 \UpArrow & 9.090 \UpArrow & 16.90 \UpArrow & 10.28 \UpArrow & 8.878 \UpArrow\\
    Recall & --9.33 \DownArrow & --10.6 \DownArrow & --10.9 \DownArrow & --7.61 \DownArrow & --4.59 \DownArrow & --5.36 \DownArrow\\
    F1-Score & 4.551 \UpArrow & --0.28 \DownArrow & --1.90 \DownArrow & 5.276 \UpArrow & 2.714 \UpArrow & 1.372 \UpArrow\\
    MCC & 4.665 \UpArrow & --0.24 \DownArrow & --1.84 \DownArrow & 5.382 \UpArrow & 2.784 \UpArrow & 1.448 \UpArrow\\
    \hline
    \end{tabular}%
  \label{tab:noise_reduction_models}%
\end{table}%

\subsection{\textbf{RQ3}: Do Noise-Reduction Techniques Improve the Performance of SV Prediction Models Using Auto-labeled SVs?}
\label{subsec:rq3_results}
 
The noise-reduction models performed similarly or better, while using much less data, compared to the original models in RQ2, as shown in Tables~\ref{tab:noise_reduction_models} and~\ref{tab:retained_rates}.
CL performed the best overall (CL $>$ CR $>$ DR), enhancing MCC by 4.7\% and 5.4\% for the D2A and combined models, respectively, significant with $p$-values of 1.6e-3 and 3.3e-3 and non-negligible effect sizes of 0.145 and 0.273, as per Wilcoxon signed-rank tests~\mbox{\cite{wilcoxon1992individual}}.
In total, combined models augmented with CL had 30\% (OpenSSL), 5\% (FFmpeg), 15.3\% (both projects) higher average MCC than HLD models in RQ2.
The domain-specific technique (DR) could not outperform the general technique (CL), showing the strong generalizability to the file-level SV prediction task of the state-of-the-art data-cleaning CL method.
However, all the noise-aware models required only 48\% -- 62\% on average and as low as 10\% of the D2A-labeled SVs to perform on par with the original models.
The similar performance with fewer samples means that many of the removed cases were indeed non-vulnerable and did not contribute to the SV predictive performance.
We further explored changing the labels of the D2A files to be removed by the noise-reduction techniques from vulnerable to non-vulnerable.
This scenario reduced the performance (MCC) of CL, CR, and DR based models by up to 2\%, 9\%, and 17\%, respectively.
These reductions suggest that some of the removed cases, i.e., deemed non-vulnerable by the noise-reduction methods, were actually vulnerable.

\begin{table}[t]
\fontsize{8.5}{9.5}\selectfont
  \centering
  \caption{Average \& smallest (in parentheses) percentages (\%) of auto-labeled SV samples retained by the noise-reduction techniques. \textbf{Notes}: Bold \& gray values are the lowest row-wise. DR is model-agnostic with constant retained \%.}
    \begin{tabular}{lccc}
    \hline
    \multirowcell{2}[0ex][l]{\textbf{Project}} & \multicolumn{3}{c}{\textbf{Noise-reduction techniques}}\\
    \cline{2-4}
    & \textbf{CL} & \textbf{CR} & \textbf{DR} \\
    \hline
    OpenSSL & \cellcolor[HTML]{C0C0C0}\textbf{36.4 (10.2)} & 47.7 (33.1) & 45.5 (45.5) \\
    \hline
    FFmpeg & \cellcolor[HTML]{C0C0C0}\textbf{60.2 (30.9)} & 76.1 (54.5) & 76.0 (76.0) \\
    \hline
    \hline
    Average of the projects & \cellcolor[HTML]{C0C0C0}\textbf{48.3 (20.6)} & 61.9 (43.8) & 60.8 (60.8) \\
    \hline
    \end{tabular}%
  \label{tab:retained_rates}%
\end{table}%

We unveiled the nature of the removed and retained samples.
We analyzed 272 random samples of the D2A-labeled vulnerable files, which were of 90\% confidence level and 10\% error~\cite{cochran2007sampling}. Among them, 136 cases (68/project) were removed and the other 136 were retained by the best CL-based model with the highest validation MCC. We found 56/68 (82\%) and 27/68 (40\%) of the vulnerable files were non-vulnerable (with no developers' mention of SV fixes) yet retained for OpenSSL and FFmpeg, respectively. Most of the probably wrongly retained cases were \textit{project-specific fixes}, as discussed in RQ1. We speculate that the patterns of these cases were rare, so the models could not effectively distinguish them.
Besides, 10/68 (15\%) and 47/68 (69\%) of the vulnerable files were incorrectly removed for OpenSSL and FFmpeg, respectively. The lower accuracy in noise removal and sample retention in FFmpeg could explain the lower performance gains for FFmpeg than OpenSSL (see \tab~\ref{tab:noise_reduction_models}).

We discovered two key reasons for the incorrectly removed cases. The first scenario involved vulnerable code coming from an external function outside of the current file, limiting the context for models to identify the SV. For example, in the commit \textit{59a56c4} of OpenSSL, the pointer *\code{p8} returned by the external function \code{EVP\_PKEY2PKCS8} outside of the file \code{crypto/asn1/i2d\_pr.c} could be \code{NULL} and lead to a null pointer dereference. However, the current file did not have sufficient information about whether a NULL check had been performed in the external function.
Another scenario of incorrect removal was with rare and context-specific SVs, making the respective patterns less frequent/evident in a training set for models to recognize. For instance, in the commit \textit{1302ccc} of FFmpeg, the value of \code{n} in the loop could cause a buffer overflow when $n \times 2 > 128$ (the maximum size of the buffer \code{synth\_pf}).
Such SV was size/buffer-dependent and thus did not appear often in the project.
These findings imply that the noise-reduction techniques not only remove (\textit{i}) noisy (incorrectly labeled) samples that help increase model confidence in predicting common SV patterns, but also (\textit{ii}) clean samples with unique/diverse patterns that can increase SV detection coverage. Such observations are supported by the significant increases in Precision yet decreases in Recall when using these techniques (see \tab~\ref{tab:noise_reduction_models}).
Overall, using auto-labeled data with human-labeled data and noise reduction produces the best SV predictive performance. We also distilled opportunities for reducing noise in auto-labeled SV data in the future.

\begin{tcolorbox}
\textbf{RQ3 Summary}. Confident Learning is the best noise-reduction method, enhancing the original (RQ2) performance by 2-30\% while using less than 50\% of the auto-labeled SVs.
Yet, there are still inaccuracies in noise reduction, especially for project-specific non-security bugs. Thus, researchers/practitioners should not rely solely on these methods to address noise in (auto-labeled) SV data.
\end{tcolorbox}

\section{Threats to Validity}\label{subsec:threats_to_validity}

There are threats to the label completeness and correctness of the ground-truth files.
We reduced the threats by augmenting NVD with security advisories that significantly increased the number of human-labeled vulnerable files.
Data from these sources are of high quality as they have been vetted and frequently maintained.
For the non-vulnerable files, we carefully removed security-related ones. We also validated the ground-truth files.

Subjectivity in manual analysis can be another threat.
Still, the manual analyses were done on significant samples by a researcher with sufficiently relevant knowledge and experience and then validated by a senior researcher with a long-term track record in the field.
We also increased the reliability of the analyses by relying on developers' explicit mentions of SV fixes.
We might miss correctly auto-labeled cases yet not human-acknowledged. 
However, these cases are extremely difficult to reliably verify as they usually involve code outside of a commit/file of interest.

The optimality of the SV prediction models may be of concern. With limited computational resources, we could not try all possible hyperparameters. We still tuned our models using the common hyperparameters from relevant studies.

The finding generalizability and reliability are also concerns. We mitigated the former using two ubiquitous C/C++ projects with mature SV reporting practices.
We also shared our data and code~\cite{reproduction_package_esem2024} for future extensions to other granularities, languages, SV tasks, and domains.
We addressed the latter by using the Wilcoxon signed rank test and its effect size to check the significance of key findings.

\section{Conclusion}
\label{sec:conclusions}

We explored the quality and usefulness of auto-labeled SVs for SV prediction, with respect to human-labeled SVs.
We first uncovered the noise level (50+\%) and the noise patterns of the large-sized auto-labeled SVs in the state-of-the-art D2A dataset.
Despite the noise, auto-labeled SVs could improve the SV predictive performance (MCC) up to 17\% when used alone and up to 22\% when used together with human-labeled SVs.
We also highlighted the benefits and challenges of automatically combating noise in auto-labeled SVs using the contemporary noise-reduction methods.
Given the rising development and use of SV auto-labeling for SV prediction, our study alerts the community to the noisy nature of the data.
We strongly recommend auto-labeled SV data should always be validated to ensure the reliability and performance of resultant SV prediction models.
We also call for more effective noise-tackling techniques for (auto-labeled) SV data to maximize their utilization for downstream SV tasks.

\section*{Acknowledgments}
The work has been supported by the Cyber Security Research Centre Limited whose activities are partially funded by the Australian Government's Cooperative Research Centres Program.

\balance

\bibliographystyle{ACM-Reference-Format}
\bibliography{reference}

\end{document}